\documentclass[%
preprint,
superscriptaddress,
showpacs,
amsmath,amssymb,
aps,prmaterials,nolongbibliography,
floatfix,
citeautoscript,
]{revtex4-2}

\usepackage[export]{adjustbox}
\usepackage[caption=false]{subfig}
\usepackage{graphicx}
\usepackage{xfrac}
\usepackage{dcolumn}
\usepackage{bm}
\usepackage{xspace}
\usepackage{textcomp}
\usepackage{siunitx}
\newcolumntype{d}[1]{D{.}{.}{#1}}
\usepackage{multirow}
\usepackage[english]{babel}

\setcitestyle{super}

\usepackage{color}

\definecolor{grey}{rgb}{.65,.65,.65}

\begin{document}

\newcommand{\mn}{$^{55}$Mn\xspace}
\newcommand{\nmg}{Ni$_2$MnGa\xspace}
\newcommand{\VV}{$V_{zz}$\xspace}
\newcommand{\cq}{$C_\mathrm{Q}$\xspace}
\newcommand{\tg}{$t_\mathrm{2g}$\xspace}
\newcommand{\xy}{$d_\mathrm{xy}$\xspace}
\newcommand{\xz}{$d_\mathrm{xz}$\xspace}
\newcommand{\yz}{$d_\mathrm{yz}$\xspace}
\newcommand{\mub}{$\mu_\mathrm{B}$\xspace}
\newcommand{\mnmn}{$\mathrm{Mn_{Mn}}$\xspace}
\newcommand{\mnni}{$\mathrm{Mn_{Ni}}$\xspace}
\newcommand{\mnga}{$\mathrm{Mn_{Ga}}$\xspace}
\newcommand{\niga}{$\mathrm{Ni_{Ga}}$\xspace}

\newcommand{\texp}[1]{\ensuremath{\times 10^{#1}}}

\DeclareSIUnit\angstrom{\text{\AA}}
\DeclareSIUnit[]\magneton{\text{\ensuremath{\mu_{\textup{B}}}}}


\title{
Structural and chemical disorder in Ni$_2$MnGa Heusler alloy investigated by nuclear magnetic resonance}

\author{Vojt\v{e}ch Chlan}
\email{vojtech.chlan@mff.cuni.cz}
\author{Martin Adamec}
\affiliation{Charles University, Faculty of Mathematics and Physics, Department of Low Temperature Physics, V Holešovičkách 2, Prague 18000, Czech Republic}
\author{Oleg Heczko}
\affiliation{Institute of Physics of the Czech Academy of Sciences,
Na Slovance 1999/2, Prague 18200, Czech Republic}

\begin{abstract}
The local environment of Mn atoms in stoichiometric Ni-Mn-Ga Heusler alloys was investigated using Nuclear Magnetic Resonance (NMR) and interpreted with the help of Density Functional Theory (DFT) methods. In cubic austenite, the significant amount of structural defects was observed in \mn NMR experiments and interpreted using DFT calculations as individual antisite defects or defects accompanying anti-phase boundaries. Combined NMR and DFT analysis provides a consistent microscopic description of local disorder in Ni-Mn-Ga.
\end{abstract}

\maketitle

\section{Introduction}
Ni-Mn-Ga Heusler alloys are ordered compounds known for their magnetic shape memory effects,\cite{Heczko2000,Straka2019,Heczko2022} one of the multiferroic effects utilizing ferromagnetism and ferroelasticity. At room temperature, the stoichiometric \nmg Heusler phase is cubic, ordered L2$_1$ structure \cite{Webster1984} called austenite. It is ferromagnetic with Curie point at about \qty{380}{K}. At about \qty{200}{K}, the parent phase transforms into a five-layered modulated phase \cite{Martynov1992} now marked 10M, called martensite.\cite{Heczko2022} This martensitic transformation, essential for multiferroicity in this material,  is thermoelastic, displacive, and diffusionless with small thermal hysteresis of several degrees. 

The martensitic transformation is driven by elastic softening of parent phase\cite{bhattacharya2003,Seiner2013} and can be strongly affected by structural disorder\cite{Seiner2013b}. Moreover, several other pecularities occur prior martensitic transformation such as spin reorientation in premartensite\cite{Perevertov2024} and non-ergodic magnetoelastic damping probably caused by the interaction of magnetic domain structure with antiphase boundaries. \cite{Bodnrov2020} After transformation the presence of antiphase boundaries in modulated martensite significantly affects its magnetic behavior and increase magnetic coercivity \cite{Straka2019} This  can provide new magnetic shape memory functionality.\cite{Straka2014} 

In ordered cubic \nmg the transition between B2 and L2$_1$ structures is assumed to be of the first order and occurs at high temperature, about \qty{1050}{K}, \cite{Overholser1999} it is thus expected to contain antisite defects and antiphase boundaries (APB). The presence of the APB boundaries in ordered Heusler alloys is well established \cite{Lapworth1974, Straka2018, Vronka2020,Venkateswaran2007}. However, the precise nature of APB region in Ni-Mn-Ga is not clear, as there is no structural TEM contrast due to the similarity of the constitute atoms.\cite{De_Graef2003} Recent calculation based on experiment suggests that APB region consists of a pair of sharp antiphase boundaries separated by single atomic layer, forming very thin APB region.\cite{Heczko2026} 

In principle, NMR can provide unique insight into the local crystal and magnetic structure of ordered Heusler alloys containing Mn. However, NMR spectra of ferromagnetic materials are sometimes difficult to interpret. We show that coupling the calculations of the electronic structure with NMR experiments provides information not available by other methods, establishing NMR spectroscopy as a method capable of revealing new important features in complex Heusler alloys. 
Here, we probe the local surrounding of Mn atoms in stoichiometric \nmg cubic austenite phases after different heat treatment and analyze structural defects such as antisites and antiphase boundaries expected in ordered compound.

\section{Methods}
Two single crystalline stoichiometric Ni-Mn-Ga samples with different thermal treatment (annealed and quenched) were used; at first the grown single crystal was subjected to homogenization at 1273 K and then ordering annealing at 1073 K and water quenched. Part of the crystal was then annealed again at 1073 K and slowly cooled in the furnace. 

The composition of the samples was determined by X-ray fluorescence (XRF) having stoichiometric composition within XRF error. Both samples were ferromagnetic, with Curie point about \qty{377}{K} and transformation to 10 M martensite below 210 K. At about 250 K the samples exhibited premartensite transformation. 
Curie and transformation temperatures were determined from the temperature dependence of low-field DC magnetization using PPMS vibrating sample magnetometer (Quantum Design) \cite{Straka2018}.  The presence of antiphase boundaries was ascertained in stoichiometric Ni-Mn-Ga austenite state  at room temperature \cite{Bodnrov2020} by magnetic force microscopy (MFM). 

\mn NMR spectra were measured without external static magnetic field at room temperature using Bruker Avance II spectrometer. Home-built probe was properly tuned and impedance-matched at each excitation frequency. All spin echoes induced by the applied Carr-Purcell-Meiboom-Gill (CPMG) multipulse sequence were coherently summed in the time domain, and their sum Fourier transformed. The resulting NMR spectrum was constructed from amplitudes of the Fourier transforms (FT) at individual excitation frequencies or constructed as an envelope of complete FTs depending on the frequency step. The excitation conditions were carefully adjusted to produce the maximum signal at a given frequency: typical lengths of applied radiofrequency pulses were \qty{0.5}{$\mu$s}, with \qty{9}{$\mu$s} delays between pulses and \qty{5}{ms} delays between subsequent scans. 4 spin echoes were recorded in each scan and 131 072 scans were accumulated to obtain a reasonable signal-to-noise ratio at room temperature.

To model the electronic structure of \nmg, we utilized density functional theory calculations using the all-electron full-potential linearized augmented plane-wave + local orbitals method as implemented in the WIEN2k software package\cite{Blaha2020}. The GGA exchange-correlation potential (Perdew-Burke-Ernzerhof form\cite{Perdew1996}) was employed with treatment of electronic correlations by the GGA+U approach applied for the Mn 3d states, using parameters $U_\mathrm{eff}=U-J=1.8$~eV and $J=0$~eV in agreement with previous calculations. \cite{Zeleny2021} The computational parameters were checked for convergence with respect to hyperfine magnetic fields; for example, in the unperturbed fcc unit cell, we used a basis set of 333 functions ($R_\mathrm{MT}K_\mathrm{max} = 8.0$) and 455 k-points in the irreducible part of the Brillouin zone (corresponding to $25\times 25\times 25$ k-mesh). The spin-orbit coupling was taken into account to obtain all components of hyperfine magnetic field.

\section{results and discussion}

\subsection{Hyperfine field in $\mathrm{Ni}_2\mathrm{MnGa}$ - considerations}

Nuclear magnetic resonance in magnetic \nmg arises due to hyperfine magnetic field $B_\mathrm{hf}$ induced at \mn nucleus by the surrounding electrons:
\begin{eqnarray}\label{eq:bhf}
 B_\mathrm{hf}&=&B^\mathrm{core}_\mathrm{Fermi} + B^\mathrm{4s}_\mathrm{Fermi} + B_\mathrm{orb} + B_\mathrm{dip}.
\end{eqnarray}
Fermi contact field, in the equation above decomposed as $B^\mathrm{core}_\mathrm{Fermi} + B^\mathrm{4s}_\mathrm{Fermi}$, originates from the contact interaction between the nuclear and electronic spin of s electrons and amounts tens of Tesla for nuclei of magnetic 3d atoms, while the orbital ($B_\mathrm{orb}$) and dipolar ($B_\mathrm{dip}$) terms are about an order of magnitude smaller ($\sim$\SIrange[range-units=single]{0.1}{1}{T}). The core contribution $B^\mathrm{core}_\mathrm{Fermi}$ reflects polarization of the 1s, 2s, and 3s electrons by on-site 3d electrons and thus scales with the Mn 3d magnetic moment.\cite{Novak2010} The valence contribution $B^\mathrm{4s}_\mathrm{Fermi}$ stems from spin-polarized 4s electrons affected by the crystal environment, hence it is highly sensitive to local atomic configuration.

In addition to the hyperfine magnetic field, the total local magnetic field at the \mn nuclei also comprises a small (\qty[parse-numbers=false]{<0.1}{T}) magnetic field due to dipolar interactions with atomic moments within the Lorentz sphere. 
The \mn nucleus (spin $I=5/2$) has an appreciable electric quadrupole moment \qty[parse-numbers=false]{Q=0.33(1)}{b} (\qty{1}{b}~$=$~\qty{e-28}{m\squared}), therefore, splitting onto five spectral lines may appear if an electric field gradient is present. In \nmg, the high site symmetry suppresses this effect, so  quadrupole splitting is expected to be weak.

DFT calculations underestimate $B_\mathrm{hf}$ for magnetic atoms due to inadequate treatment of core-valence exchange polarization; calculated $B^\mathrm{core}_\mathrm{Fermi}$ is typically lower by tens of percent.\cite{Novak2010,Blugel1987} In austenite \nmg, the calculated hyperfine field $B^\mathrm{core}_\mathrm{Fermi}$ = \qty{-47.2}{T} (minus sign denotes the antiparallel orientation with respect to the magnetic moment of Mn atom), $B^\mathrm{4s}_\mathrm{Fermi}$=\qty{31.8}{T}, and $B_\mathrm{orb}\!+\!B_\mathrm{orb}\!=\,$\qty{0.25}{T}. In NMR experiments\cite{Belesi2015}, the \mn resonance near $\qty{305}{MHz}$ at low temperatures corresponds to $B_\mathrm{hf}$ =\qty{28.8}{T}, resulting in $B^\mathrm{core}_\mathrm{Fermi} \sim \qty{-60.9}{T}$, which is underestimated by approximately \qty{23}{\%} in calculations. Despite this, the calculated fields remain useful for interpreting NMR trends--in particular relative shifts, their signs, and the number of expected spectral lines--which is utilized in the analysis below.

\subsection{Defects in Austenite $\mathrm{Ni}_2\mathrm{MnGa}$}

\begin{figure}
	\centering
	\includegraphics[width=\columnwidth]{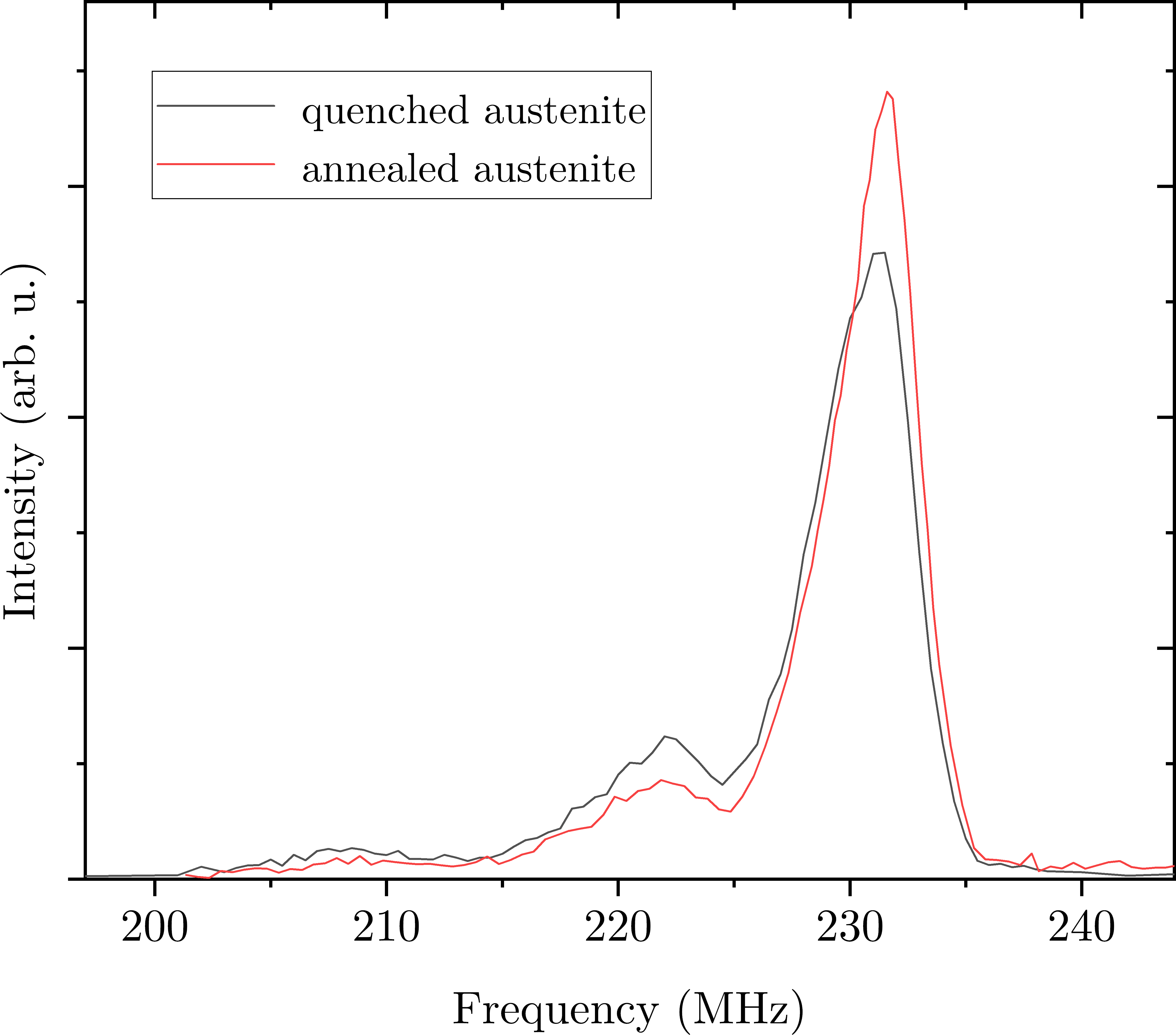}
	\caption{\mn NMR spectrum of quenched and annealed \nmg samples obtained in zero external magnetic field at room temperature.} 
	\label{f1:Aspect}
\end{figure}

In ordered structures various structural and chemical imperfections can be present. However, in Ni-Mn-Ga cubic compound no dislocations were observed, there is no twinning and single crystal excludes the grain boundaries. Thus, the only defects which can be present are related to the ordering. These are vacancies, antisites, and antiphase boundaries.

In austenitic cubic phase \nmg with L2$_1$ symmetry, all Mn atoms are equivalent and occupy cubic sites (point group $m3m$). The electric quadrupole interaction vanishes, so a single resonance line is expected, determined by $B_\mathrm{hf}$ (Eq.~\ref{eq:bhf}).
In contrast the \mn NMR spectrum of the stoichiometric \nmg measured at room temperature (Fig.~\ref{f1:Aspect}) displays one intensive resonance line and two weaker satellite lines. The position of the main resonance line at \qty{231}{MHz} coincides with that reported in previous studies\cite{OConnor2002,Belesi2015}, and corresponds to $|B_\mathrm{hf}|\sim\qty{21.8}{T}$ ($f=\gamma_\mathrm{Mn}B_\mathrm{hf}$, $\gamma_\mathrm{Mn}=\qty{10.576}{\MHz\per\tesla}$). Furthermore, two weaker lines at lower frequency (208 and \qty{222}{MHz}) are well resolved in the \mn NMR spectrum. In a previous study\cite{Belesi2015}, these weaker lines were not fully resolved and instead a tail of the main line toward lower frequencies was observed. This spectral shape was attributed to a slight off-stoichiometry in the \nmg austenite samples. For the stoichiometric samples presented in this study, we attribute the well resolved weaker satellite lines to structural defects related to a structural and chemical disorder.

The hyperfine magnetic field is highly sensitive to changes in a local electronic structure especially via its term $B^\mathrm{4s}_\mathrm{Fermi}$, and thus the resonance frequency of the \mn nuclei in the vicinity of a structural defect is modified and these \mn nuclei resonate at different frequencies compared to nuclei more distant from such defect. In \nmg the the weaker lines observed at 208 and \qty{222}{MHz} can be ascribed to several different antisite defects\cite{Malik2022,Salvador2022} and anti-phase boundaries\cite{Heczko2020,Venkateswaran2007}. The concentration of these structural defects can be reduced by thermal treatment, which is reflected in comparison of the \mn NMR spectra of the annealed and quenched \nmg sample (Fig.~\ref{f1:Aspect}): the main line at \qty{231}{MHz} became slightly narrower for the annealed sample compared to the quenched sample, and the intensity of the satellite lines decreased with annealing, indicating a decreased amount of defects. However, the type of chemical disorder or structural defect cannot be unambiguously assigned by the NMR experiment alone.

\begin{figure}
	\centering
	\includegraphics[width=\columnwidth]{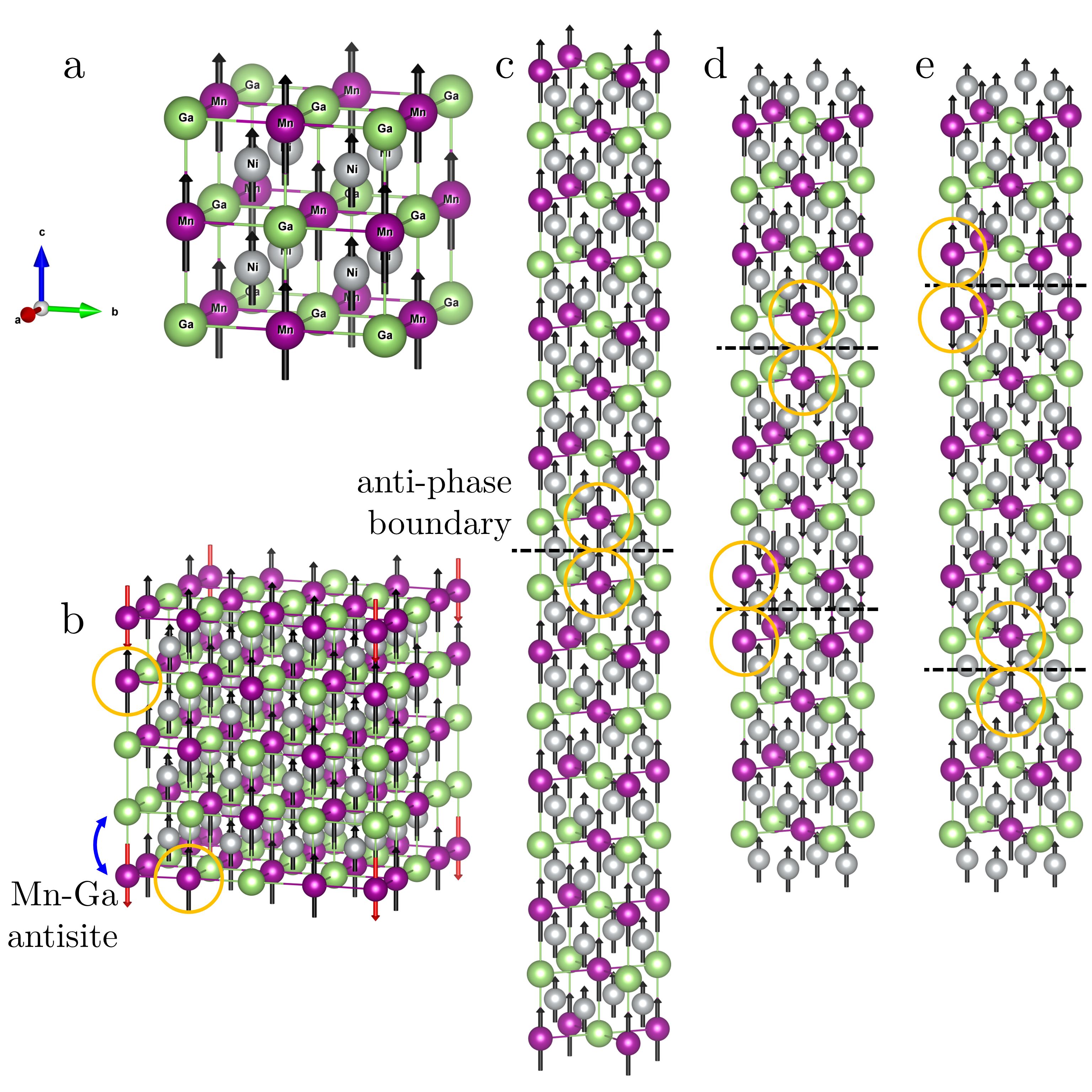}
	\caption{Calculated austenite structures of \nmg. (a) Ideal cell, (b) Mn$\leftrightarrow$Ga antisite (blue arrow), (c) one APB with ferromagnetic order, (d,e) two APBs with antiferromagnetic order. Red and yellow arrows highlight Mn orientations and nearest Mn neighbors.}
	\label{f2:Astr}
\end{figure}

To identify likely defects, we calculated several DFT models and evaluated parameters relevant to NMR experiment: fully ordered cubic austenite, and $2{\times}2{\times}2$ supercells with Mn$\leftrightarrow$Ga, Mn$\leftrightarrow$Ni, or Ga$\leftrightarrow$Ni antisite. Given the B2$\to$L2$_1$ transition at high temperature, Mn$\leftrightarrow$Ga is expected to be the most probable, while exchange with Ni should play significantly smaller role as A2 disordered phase is now observed even in high temperature. We also modeled single and two-boundaries APB structures where a phase shift was introduced into the sequence of alternating stacking of Mn and Ga (Fig.~\ref{f2:Astr}). The two-boundary APB with narrow core of one atomic layer was recently proposed based on modeling of  thermal antiphase boundary in \nmg.\cite{zemen2025}. This structure is energetically favorable compared to sharp single antiphase boundary. Still the precise atomic-scale structure of APBs remains uncertain since they are observed only magnetically\cite{Straka2018,Heczko2020,Venkateswaran2007,Vronka2020}, and the boundary can be formed from more than one atomic layer as observed for thermally induced interface.\cite{Murakami2014}

Introducing antisites splits the originally equivalent Mn atoms into 10–12 non-equivalent sites. The separation of defects within the supercells is large enough so that distant Mn retain values of $B_\mathrm{hf}$ unchanged from the unperturbed austenite unit cell. Only Mn near defects show notable hyperfine-field shifts, which should give rise to observed satellite lines. Figure~\ref{f2:Astr} displays all considered structural variants and Table~\ref{tab:calc} summarizes the calculated frequency shifts. In the cases of Mn and Ni atoms participating in the antisite, both spin orientations were considered, and the one with the lowest total energy was used. For Mn in Ga sublattice (\mnni), the calculated Mn moments couple antiferromagnetically to bulk Mn,  which is in agreement with the experimental observations of the dependence of the total magnetic moment on deviation from stoichiometry \cite{Enkovaara2003} and the general tendency in Heusler alloys\cite{Lapworth1974}. Similar antiferromagnetic scenario is realized for Mn in Ni sublattice (\mnni), the calculated Mn moments also prefer an antiferromagnetic orientation to that of the Mn sublattice.  For the Ga$\leftrightarrow$Ni case, the calculated moment of Ni placed within the Ga sublattice (\niga) couples ferromagnetically to the Mn sublattice, even when initialized oppositely.

\begin{table}[htbp]
  \centering
  \caption{Calculated resonance frequency shifts of \mn nuclei at antisite and antiphase boundaries (denoted as defect), their nearest \mn neighbors ("n.n"), and more distant ones ("bulk"), the shift is referred to the frequency of unperturbed \nmg austenite. The differences of total energies per $1{\times}1{\times}1$ unit cell and with respect to unperturbed \nmg austenite structure are displayed in the last column, indicating the most preferred perturbed structures. }
  \begin{ruledtabular}
    \begin{tabular}{ld{2}d{0}d{0}d{0}}
   \multirow{2}{*}{structure} & \multicolumn{3}{c}{$\Delta f$(MHz)} &  \multicolumn{1}{c}{\multirow{2}{*}{$\Delta E$ (meV)}} \\
              & \text{defect} & \text{n.n}.    & \text{bulk}  &  \\
              \hline 
    Mn$\leftrightarrow$Ga antistite & -72 & -17 & \numrange[range-phrase = {\text{ to }}]{-1}{1}  &55 \\
    Mn$\leftrightarrow$Ni antistite & -110 & -45 & \numrange[range-phrase = {\text{ to }}]{-1}{1}{}   &143\\
    Ga$\leftrightarrow$Ni antistite & \multicolumn{1}{c}{--} & -10 &\numrange[range-phrase = {\text{ to }}]{-2}{2}{}   &215 \\
        APB1 & 27  & -3    & \numrange[range-phrase = {\text{ to }}]{-2}{-1}{}  &53 \\
    APB2 & -25  &  2   & \sim-1  &31 \\
    APB2c & -24 & 4   & \numrange[range-phrase = {\text{ to }}]{-1}{0}{} &30 \\
    \end{tabular}
    \end{ruledtabular}
     \label{tab:calc}
\end{table}

Calculated magnetic moments and hyperfine fields for structures with defects reveal that the impact of a defect is very localized: essentially only the atoms participating in the defect and their nearest neighbors are noticeably affected, whereas the properties of more distant ("bulk-like") atoms farther than \qty{5}{\AA} already coincide with the values of the pure \nmg structure.

An example of how the calculated data in Table~\ref{tab:calc} can help to interpret the NMR spectra is demonstrated on Mn$\leftrightarrow$Ga case. The magnetic moments of the "bulk-like" Mn$_\mathrm{Mn}$ atoms in Mn$\leftrightarrow$Ga structure are equal to the value of \qty{3.79(2)}{\magneton} of Mn moments in pure \nmg (with an estimated accuracy margin of \qty{0.02}{\magneton}). Consequently, also the hyperfine fields of the "bulk-like" Mn atoms coincide with the value in pure \nmg within the precision estimate of the calculations $\sim\qty{0.2}{T}$, which gives essentially zero shift in frequency (see Table~\ref{tab:calc}). On the other hand, the calculated moment of Mn$_\mathrm{Ga}$ involved in the antisite is \qty{-3.93(2)}{\magneton}, which leads to a slight increase in $B^\mathrm{core}_\mathrm{Fermi}$ by \qty{1.1(5)}{T}. This shift is overcome by the opposing valence contribution $B^\mathrm{4s}_\mathrm{Fermi}$, which for Mn$_\mathrm{Ga}$ is by \qty{7.9(5)}{T} larger than for Mn$_\mathrm{Mn}$ in pure \nmg, thus reducing the total hyperfine field. Therefore, the calculations predict that \mn resonance of Mn$_\mathrm{Ga}$ should be significantly decreased (\qty{\sim72}{MHz}) relative to the bulk Mn$_\mathrm{Mn}$; differences in $B_\mathrm{orb}$ and $B_\mathrm{dip}$ are negligible.

Such a large shift arises as a result of completely different crystal environments and exchange interactions: Mn$_\mathrm{Ga}$ is antiferromagnetically coupled to 8 neighboring Ni atoms (in the first shell) and to 5 Mn + 1 Ga (in the second shell), while unperturbed Mn$_\mathrm{Mn}$ is ferromagnetically coupled to the 8 neighboring Ni atoms and then surrounded by 6 Ga atoms. A similar but weaker impact can be observed for the five Mn atoms closest to the antisite defect: for these Mn one of the six Ga neighbors was replaced by the antisite Mn$_\mathrm{Ga}$, which leads to the calculated shift of \qty{-17}{MHz} (Table~\ref{tab:calc} column n.n.). The significantly disturbed electronic structure of Mn$_\mathrm{Ga}$ causes only about 4\texttimes~larger shift compared to the nearest Mn$_\mathrm{Mn}$, where the second shell is perturbed much less. This is somewhat surprising, since the contributions of each exchange-interaction partner to the valence term $B^\mathrm{4s}_\mathrm{Fermi}$ are often nearly additive. We suppose that the transferred hyperfine field from neighboring Mn atoms is partially compensated for by the antiferromagnetic coupling of Mn$_\mathrm{Ga}$ to the Ni sublattice. 

Our calculations predict negative shifts of the resonance frequencies of \mn near a defect for all structures with antisite defects (Table~\ref{tab:calc}), that is, in general the antisite defect tends to decrease the resonance frequency of neighboring \mn. The antisite defects can be expected in the ordered structure, however, no direct proof exists. 

On the other hand the structural defects which have been experimentally observed in Ni-Mn-Ga \cite{Straka2018, Vronka2020, Heczko2026} are thermally induced antiphase boundaries arising from B2$\to$L2$_1$ ordering transformation. Recent calculation and observation suggested that the antiphase region consists of two sharp antiphase boundaries with thin core of single atomic layer.\cite{zemen2025, Heczko2026}  Such double structure exhibits a similar effect in NMR as antisite defects, where \mn at the boundary is mostly affected and the effect decreases rapidly with increasing distance.  

In contrast, the structure with a single antiphase boundary (APB1) behaves differently and \mn atoms that form the boundary have positive frequency shift (Table~\ref{tab:calc}). In the single antiphase boundary the local environment of resonating \mn is almost identical to that of Mn$_\mathrm{Mn}$ n.n.\ in Mn$\leftrightarrow$Ga, i.e., one Mn and five Ga in the second shell, but there is a difference in the orientation of the magnetic moment of this neighboring Mn, which induces an opposite contribution to the hyperfine field $B^\mathrm{4s}_\mathrm{Fermi}$. In Mn$\leftrightarrow$Ga the antisite Mn$_\mathrm{Ga}$ is antiferromagnetically coupled to its n.n.\  Mn$_\mathrm{Mn}$, in the case of an antiphase boundary, the Mn atoms that form the defect have their moments parallel. In addition, the healing length of APB is relatively short: already the nearest Mn neighbors have their hyperfine fields very close to the bulk-like \mn. The difference between single APB and thick APB region separated by two APBs is significant for the interpretation of NMR spectra as positive frequency shift was not observed in the NMR spectra. Moreover, the calculation demonstrated that the single boundary exhibit large excess energy. \cite{zemen2025}  

When evaluating the most probable type of defect in \nmg, we should take into account the number of predicted spectral lines, their intensities and approximate frequency shifts, as well as the calculated total energies and experimental observation of APBs. Taking into account the calculated total energies, the relevant structures are those with two antiphase boundaries and the one with the Mn$\leftrightarrow$Ga antisite, in agreement with other experiments \cite{Lazpita2011,Enkovaara2003,Richard2007}. The frequency shifts observed in the \mn spectra are close to those predicted by the calculations for APB with two boundaries and smaller than the shifts calculated for antisite defects. On the other hand, the ratio of intensities of the two satellite lines in the NMR spectrum is roughly 1:5 and thus corresponds well to the situation in Mn$\leftrightarrow$Ga or Mn$\leftrightarrow$Ni where there are 5 Mn neighbors at a distance of \SI{2.87}{Angstrom} from the one Mn that forms the antisite defect. In case of APB the ratio of lines would be 1:1, since there is one layer of Mn participating in the APB itself accompanied by one layer of nearest Mn.  

 We should also consider the absolute concentrations of the defects and here we should distinguish between annealed and quenched sample. Following the interpretation of satellite lines in the \mn spectrum as the Mn$\leftrightarrow$Ga antisite and its nearest neighbor, the antisite concentration would be approximately \qty{4}{percent} and \qty{2}{percent} for the quenched and annealed \nmg sample, respectively. Assuming that the satellite lines arise only due to APB, the average distance between two boundaries in order to reach the observed intensity of  \qty{4}{percent}  would be approximately \SI{300}{\angstrom}. For annealed sample estimated density of APB is about \SIrange[]{0.5}{2}{\per\micro\meter}. \cite{Vronka2020} Considering the small thickness of the boundary the volume of APB region is negligible and cannot be detected. Therefore, we can conclude that the lines in annealed sample are most likely induced by Mn$\leftrightarrow$Ga antisites, which is in agreement with the situation in Ni$_{2-x}$Pt$_x$MnGa\cite{Singh2016}. Antiphase boundaries are most likely also present, but their concentration is too low to be resolved in the \mn NMR spectrum.
  On the contrary in quenched sample the concentration of APBs sharply increases to \SIrange[]{15}{20} per {$\mu m$}. \cite{Vronka2020} This high concentration provides a significant volume of APB region (several percents) which should be detectable. Thus the increased intensity of the satellite lines in quenched sample can be ascribed to increasing volume of antiphase region. On the other hand, small detected volume can exclude significantly thicker APB regions as reported for other ordered alloys. \cite{Murakami2014}


\subsection{Conclusions}
 We demonstrate that NMR experiments allow to observe and, in combination with DFT calculations, to identify fine details of the structure and ordering of the \nmg Heusler alloy. The presence of Mn-Ga antisite defects is indicated in the NMR spectrum by additional weak lines at lower frequencies. In stoichiometric \nmg cubic austenite there are about 4 per cent misplaced Mn atoms in the form of antisite Ga-Mn and/or of a narrow thermal antiphase region consisting of two antiphase boundaries and a thin disordered core. This amount of defects decreases with annealing, which can be ascribed to a sharp decrease of the density of the antiphase boundaries.  In summary, NMR spectroscopy together with the electronic structure calculations provides a powerful tool for characterizing in detail the structural defects in ordered Ni-Mn-Ga magnetic shape-memory alloys.

\begin{acknowledgments}
This work was supported by the Czech Science Foundation (project No. 23-04806S) and by the Ferroic Multifunctionalities project (FerrMion)  [Project No. CZ.02.01.01/00/22\_008/0004591], supported by the Ministry of Education, Youth and Sports of the Czech Republic (MEYS) and co-funded by the European Union. Computational resources were provided by the e-INFRA CZ project (ID:90254) supported by MEYS.

\end{acknowledgments}

\bibliographystyle{apsrev4-2}
\bibliography{nimnga}

\end{document}